\documentclass[pdftex,twocolumn,epjc3]{svjour3}          

\RequirePackage[T1]{fontenc}

\smartqed  

\RequirePackage{graphicx}
\RequirePackage{mathptmx}      
\RequirePackage{flushend}
\RequirePackage[numbers,sort&compress]{natbib}
\RequirePackage[colorlinks,citecolor=blue,urlcolor=blue,linkcolor=blue]{hyperref}

\journalname{Eur. Phys. J. C}

\usepackage{amsmath}
\usepackage{xspace}
\usepackage{xcolor}
\usepackage{mathtools}
\usepackage{microtype}
\usepackage{dblfloatfix} 
\usepackage{doi}
\usepackage[switch]{lineno} 

\def\u{$^{238}$U\xspace}

\def\rnddd{$^{222}$Rn\xspace}
\def\poduo{$^{218}$Po\xspace}
\def\pbduq{$^{214}$Pb\xspace}
\def\biduq{$^{214}$Bi\xspace}
\def\poduq{$^{214}$Po\xspace}

\def\th{$^{232}$Th\xspace}
\def\raddq{$^{224}$Ra\xspace}
\def\rnddz{$^{220}$Rn\xspace}
\def\podus{$^{216}$Po\xspace}

\def\bidud{$^{212}$Bi\xspace}
\def\podud{$^{212}$Po\xspace}

\def\udtc{$^{235}$U\xspace}
\def\raddt{$^{223}$Ra\xspace}
\def\rndun{$^{219}$Rn\xspace}
\def\poduc{$^{215}$Po\xspace}

\begin{document}

\title{Improving radioactive contaminant identification through the analysis of delayed coincidences with an $\alpha$-spectrometer}

\author{
G.~Baccolo\thanksref{INFNMiB,MIB-SM}
\and A.~Barresi\thanksref{INFNMiB,MIB}
\and M.~Beretta\thanksref{INFNMiB,MIB,berkeley}
\and D.~Chiesa\thanksref{INFNMiB,MIB,e1}
\and M.~Nastasi\thanksref{INFNMiB,MIB,e2}
\and L.~Pagnanini\thanksref{LNGS}
\and S.~Pozzi\thanksref{INFNMiB,MIB}
\and E.~Previtali\thanksref{INFNMiB,MIB,LNGS}
\and M.~Sisti\thanksref{INFNMiB}
\and G.~Terragni\thanksref{MIB}
}

\thankstext{e1}{e-mail: davide.chiesa@mib.infn.it}
\thankstext{e2}{e-mail: massimiliano.nastasi@unimib.it}

\institute{
INFN - Sezione di Milano Bicocca, Milano I-20126 - Italy\label{INFNMiB}
\and
Dipartimento di Dipartimento di Scienze dell’Ambiente e della Terra, Universit\`{a} di Milano - Bicocca, Milano I-20126 - Italy\label{MIB-SM}
\and
Dipartimento di Fisica, Universit\`{a} di Milano - Bicocca, Milano I-20126 - Italy\label{MIB}
\and
INFN - Laboratori Nazionali del Gran Sasso, Assergi (L'Aquila) I-67100 - Italy\label{LNGS}
\and
\emph{Present Address}: Physics Department, University of California, Berkeley, CA 94720, USA\label{berkeley}
}

\date{Received: date / Accepted: date}

\maketitle

\begin{abstract}
In the framework of rare event searches, the identification of radioactive contaminants in ultra-pure samples is a challenging task, because the signal is often at the same level of the instrumental background. 
This is a rather common situation for $\alpha$-spectrometers and other detectors used for low-activity measurements. 
In order to obtain the target sensitivity without extending the data taking live-time, analysis strategies that highlight the presence of the signal sought should be developed. 
In this paper, we show how to improve the contaminant tagging capability relying on the time-correlation of radioactive decay sequences. 
We validate the proposed technique by measuring the impurity level of both contaminated and ultra-pure copper samples, demonstrating the potential of this analysis tool in disentangling different background sources and providing an effective way to mitigate their impact in rare event searches.

\end{abstract}

\section{Introduction}\label{intro}
Experiments searching for rare events in the field of neutrino physics and heavy dark matter searches require detectors with large active mass (up to several kton) surrounded by proper shields~\cite{Dolinski:2019nrj,Tanabashi:2018oca}.
In this scenario, the radiopurity of the materials, active or passive, is a fundamental parameter that impacts on the sensitivity and the scientific potential of the experiments themselves.
Hence the request to develop high-performance techniques and methods for material screening, allowing to set stringent limits on contaminant concentrations.

The best results in this sector are obtained from mass spectroscopy \cite{Kaizer:2019rfw,Povinec:2018wgd} and neutron activation analysis~\cite{Clemenza:2018mry}. 
These techniques measure very effectively the concentration of stable or very long half-lived nuclides (e.g. \th and \u).
By measuring a few grams of material for less than a month, a sensitivity of $\mathcal{O}(0.01)\,\mu$Bq/kg can be obtained for \th and \u, which represent the most worrisome contaminants for low background experiments.
Due to the possible breakdown of the secular equilibrium, this information is not enough to measure the activity of isotopes belonging to the lower part of the radioactive chains which produces background in many experiments~ \cite{Azzolini:2019nmi,Alduino:2016vtd}.
In this case, we exploit complementary measurements based on $\alpha$- and $\gamma$-spectroscopy, which however reach limits of 10-100 $\mu$Bq/kg \cite{Laubenstein:2020rbe}.

Among these techniques, $\alpha$-spectroscopy has some advantages that make it suitable for future developments that improve its sensitivity. 
Being sensitive both to the progenitors and to the lower part of radioactive chains (also to $\beta$-signals), $\alpha$-spectrometers provide a more complete picture of the material contaminations. 
The short $\alpha$-particle interaction range ($\sim 10~\mu$m Cu, E$_{\alpha}$\;$\sim6\;$MeV) allows for studying also the contamination diffusion through the sample, performing interesting studies on the implantation mechanism. 
The sensitivity of commercial silicon detectors for $\alpha$-spectroscopy is currently limited by their active surface ($\sim$10 cm$^2$), which can be suitably increased in the case of custom detectors. However, measurements of extremely clean samples would still be limited by the inability to separate the very small signal from the background fluctuations. 

A very effective way to identify and measure the contaminations of the lower parts of the \th, \u and \udtc natural chains consists in analyzing the time-delayed coincidences produced by decay cascades occurring in relatively short time scales. 
This technique consists in searching for couples of parent--daughter events associated with the decay of the same precursor isotope.
By exploiting the information on time correlations, it is possible to suppress the background and to bring out the distinctive signatures of different radioisotopes, especially when the complementary information provided by the energy spectrum does not allow for a clear identification of the contaminants~\cite{Chiesa:2021}.

Depending on the features of the measurement system, it is possible to detect different cascades in the natural decay chains.
The factor that plays the most important role is the time response of the detector in relation to the counting rate conditions. Obviously, the longer the half-life of the daughter nuclide, the lower must be the event rate on the detector, otherwise random coincidences would spoil the effectiveness of this technique.

For instance, bolometric detectors used in rare event searches~\cite{Adams:2021rbc,Azzolini:2019tta,Armengaud:2020luj,Abdelhameed:2019hmk,Alenkov:2019jis,Amaral:2020ryn,Arnaud:2020svb} allow to detect time coincidences up to a few minutes delay~\cite{Azzolini:2019nmi}, thanks to the very low counting rate characterizing these experiments.
 However, since they have signal rise times of the order of few ms, they are not able to resolve in time the fast Bi$-$Po sequences in the \th and \u chains. 
On the contrary, devices such as scintillating and semiconductor detectors allow to detect also faster sequences~\cite{Barabash:2017hst}, with the advantage of being able to exploit this technique also in presence of a higher counting rate.

Ion implanted silicon charged particle radiation detectors are produced with thin windows optimized for $\alpha$-spectroscopy measurements. Nevertheless they can also detect $\beta$-radiation. 
They are manufactured with surfaces from 25 to 3000~mm$^2$, thus allowing to achieve a relatively high detection efficiency in the study of surface contaminations of various samples, ranging from environmental to radiation protection or radio-assay applications.
The time resolution of these detectors is $\mathcal{O}(100)$ns, so they are suitable to study several decay sequences in the \th, \u and \udtc chains.
Moreover, they can be assembled by selecting high purity materials, thus enabling for low background measurements. Thanks to the fact that their active layer is very thin, of the order of a few hundreds $\mu m$, the muon background is negligible.
Last but not least, these devices can be easily purchased at an affordable cost.
Thanks to the aforementioned features, $\alpha$ silicon detectors are an excellent tool to search for $\alpha-\alpha$ and $\beta-\alpha$ time-delayed coincidences.

In this work, we report the results obtained with an $\alpha$-spectrometer specially adapted for the analysis of delayed-coincidences.
Particularly, we will show the results obtained from the analysis of the following decay cascades:

\begin{equation}
^{224}\text{Ra}\xRightarrow[\alpha~ (3.66~\text{d})]{5.79~\text{MeV}}~^{220}\text{Rn}\xRightarrow[\alpha~ (55.6~\text{s})]{6.40~\text{MeV}}~^{216}\text{Po}\xRightarrow[\alpha~(145~\text{ms})]{6.91~\text{MeV}}~^{212}\text{Pb}
    \label{chain:1}
\end{equation} 
\begin{equation}
\begin{split}
^{222}\text{Rn}&\xRightarrow[\alpha~ (3.82~\text{d})]{5.59~\text{MeV}}~^{218}\text{Po}\xRightarrow[\alpha~(3.1~\text{min})]{6.12~\text{MeV}}~^{214}\text{Pb}\\
^{214}\text{Bi}&\xRightarrow[\beta~(19.9~\text{m})]{3.27~\text{MeV}}~^{214}\text{Po}\xRightarrow[\alpha~(164.3~\mu\text{s})]{7.83~\text{MeV}}~^{210}\text{Pb}
\end{split}
    \label{chain:2}
\end{equation} 
\begin{equation}
^{223}\text{Ra}\xRightarrow[\alpha~(11.4~\text{d})]{5.98~\text{MeV}}~^{219}\text{Rn}\xRightarrow[\alpha~(3.96~\text{s})]{6.95~\text{MeV}}~^{215}\text{Po}\xRightarrow[\alpha~(1.78~\text{ms})]{7.53~\text{MeV}}~^{211}\text{Pb}
    \label{chain:3}
\end{equation} 
belonging to the \th (\ref{chain:1}), \u (\ref{chain:2}) and \udtc (\ref{chain:3}) decay chains, respectively.

In Sect.~\ref{Sec:setup}, we introduce the experimental setup, the tagging of delayed coincidences and the Monte Carlo reconstruction. 
In Sect.~\ref{Sec:measurements}, we detail the validation of the analysis strategy on high-contaminated sample measurements. 
In Sect.~\ref{Sec:low} finally, we proof the very-high potential of this technique investigating the contaminations of an ultra-pure copper sample.

\section{Experimental set-up and Monte Carlo reconstruction}\label{Sec:setup}

The measurements presented in this paper have been performed using an Ortec $\alpha$ detection system, which is an integrated spectrometer for measuring low activity samples. 
The detector is a ion implanted silicon semiconductor inside a vacuum chamber and is equipped with a complete amplifying chain composed by a charge sensitive pre-amplifier and an amplifier with $1~\mu$s shaping time.
Our silicon detector belongs to the Ortec series of the Ultra partially depleted detectors.
Its active area is 900~mm$^2$ and the minimum depletion depth is 100~$\mu$m. The energy resolution, measured with an $^{241}$Am $\alpha$ source,  is $35\pm5$~keV at 5.5 MeV.

The signals from the electronic chain are digitized with a CAEN multi-channel analyzer module (MCA N957) that usually only records the pulse amplitude and builds a spectrum with a resolution of 13 bit.
In order to analyze the delayed coincidences, we modified the data acquisition system to also record the timestamp of each event.
The timer frequency of this module is 10$^3$~Hz and, thus, the time resolution of our experimental set-up is 1~ms.

\begin{figure*}[!b]
    \centering
    \includegraphics[width=0.49\textwidth]{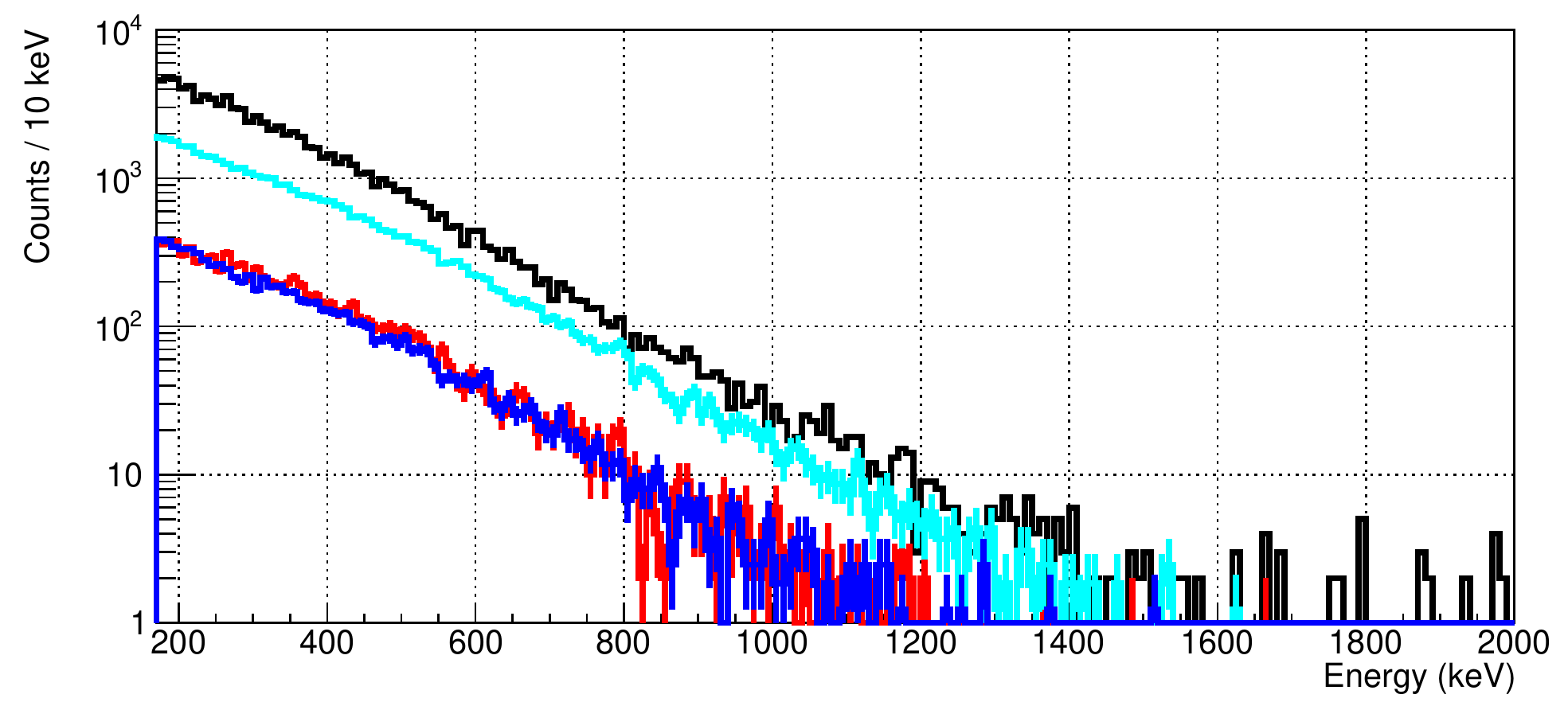}
    \includegraphics[width=0.49\textwidth]{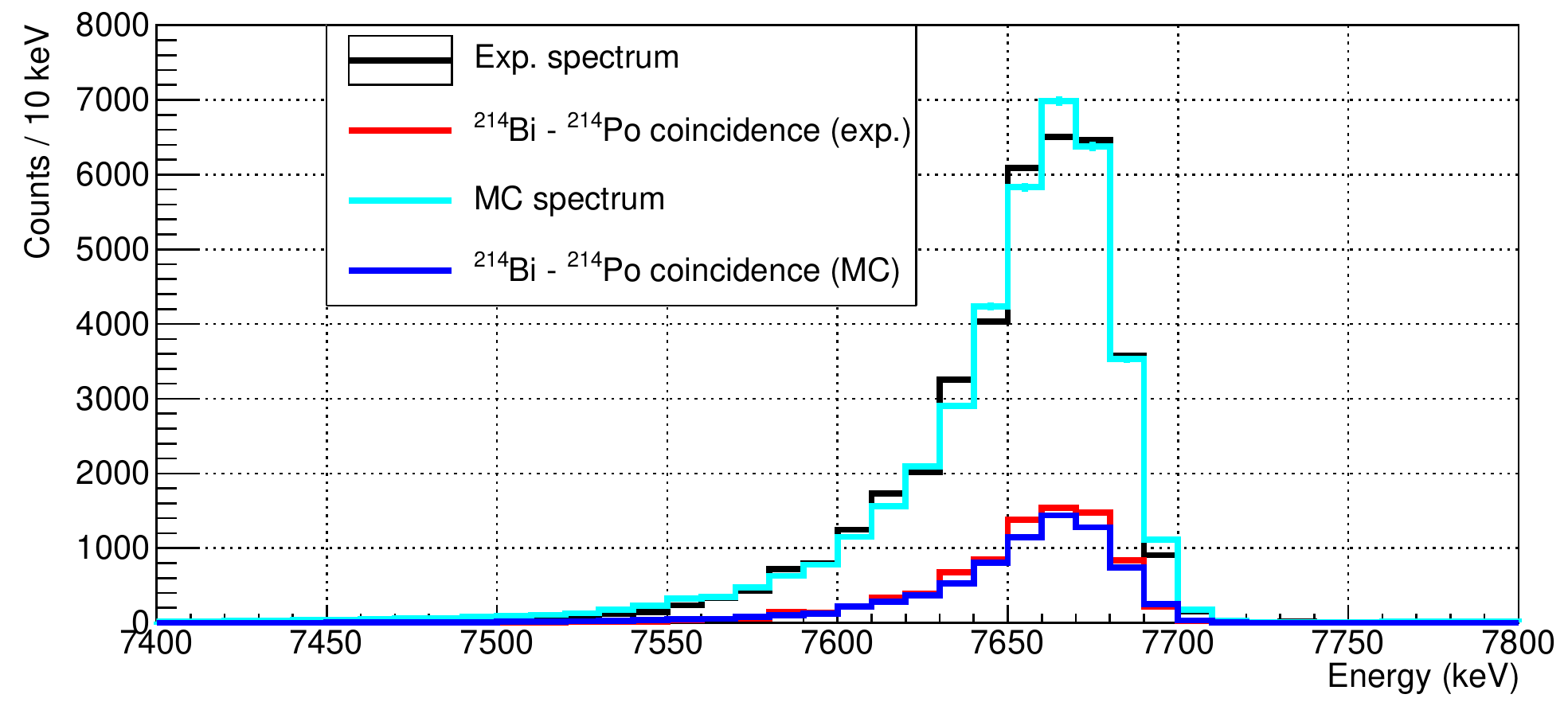}
    \caption{Experimental data (black) and MC simulations (cyan) referring to the measurement of the copper plate contaminated with the $^{222}$Rn daughters belonging to the \u chain. In the left pad we show the lower energy range where we searched for \biduq $\beta$-events in time coincidence with \poduq $\alpha$-events (right pad). The MC spectrum, comprised of \biduq and \poduq events only, is normalized at the \poduq $\alpha$ peak and, as expected, does not match the experimental data at low energies because of contributions from other $\beta$-emitters and environmental background. By selecting \biduq$-$\poduq delayed coincidences in both experimental (red) and MC (blue) data, we get a good reconstruction of both \poduq $\alpha$-peak and \biduq $\beta$-spectrum.}
    \label{fig:Spectrum_U238}
\end{figure*}

After the measurement, we analyze the raw experimental data to search for couples of parent--daughter events, like those listed at (\ref{chain:1}), (\ref{chain:2}), and (\ref{chain:3}). 
Particularly, we select the events with an energy compatible with a possible parent decay, and we open a time window ($\Delta t_w$) of the order of a few half-lives of the daughter nucleus to search for its signature.
When such a time-delayed coincidence is found, we tag both events for subsequent analyses in which the information about time correlation is exploited to improve the signal to background ratio.

In order to reproduce the experimental energy spectra and to determine the corresponding detection efficiency, we use a Monte Carlo (MC) code based on the Geant4 toolkit~\cite{GEANT}.
We implement the experimental configuration of the silicon detector starting from the factory specifications which quote an active layer in the 100$-$500~$\mu$m range and a dead layer of 50~nm.
Then, by analyzing the spectrum produced by a $\beta$ decay (see the next section~\ref{Sec:measurements}), we find that the active layer allowing for an optimal reconstruction of the measured spectral shape is $400 \pm 50$~$\mu$m.
The MC tool allows us to choose the contamination depth of the source and to simulate any natural decay chain.
The MC output is processed to introduce the detector features in terms of time and energy resolution and to record, for each event, the energy deposited in the silicon active layer and the time at which the interaction occurs.

\section{Experimental measurements with radioactive sources}
\label{Sec:measurements}

In this section, we present three measurements performed with different radioactive sources in order to test and validate the selection technique and the data analysis of delayed coincidences. 
The three sources have been produced by contaminating copper samples with different mechanisms so as to include a part or the full decay chains of \u, \udtc, and \th, respectively.
Depending on the contamination process, we obtained sources with radioisotopes distributed on the sample surface at different average depths.
In the following, we show the potential of the analysis of delayed coincidences in extracting information about surface contaminations even when the spectrum does not exhibit any peak due to the energy degradation process of escaping $\alpha$-particles from deeper surface layers.

\subsection{\textbf{Analysis of \biduq$-$\poduq delayed coincidence (\u chain)}}
\label{sec:uranium}

The goal of the first test is to identify and study the \biduq$-$\poduq delayed coincidences in the \u chain.
The source used for this measurement is a $5\times5\times0.1$~cm$^{3}$ copper plate contaminated with $^{222}$Rn daughters. 
This source was produced in a radon-box, an environment where the concentration of this gas reaches high values ($\sim$250~kBq/cm$^3$) and the $\alpha$-decaying $^{222}$Rn daughters are implanted on the sample surfaces.
After the copper plate has been exposed for 8 months in the radon-box, we started a measurement immediately after its extraction using the experimental set-up previously described for a total time of 19 hours and 20 minutes. 

\begin{figure}[t]
    \centering
    \includegraphics[width=0.49\textwidth]{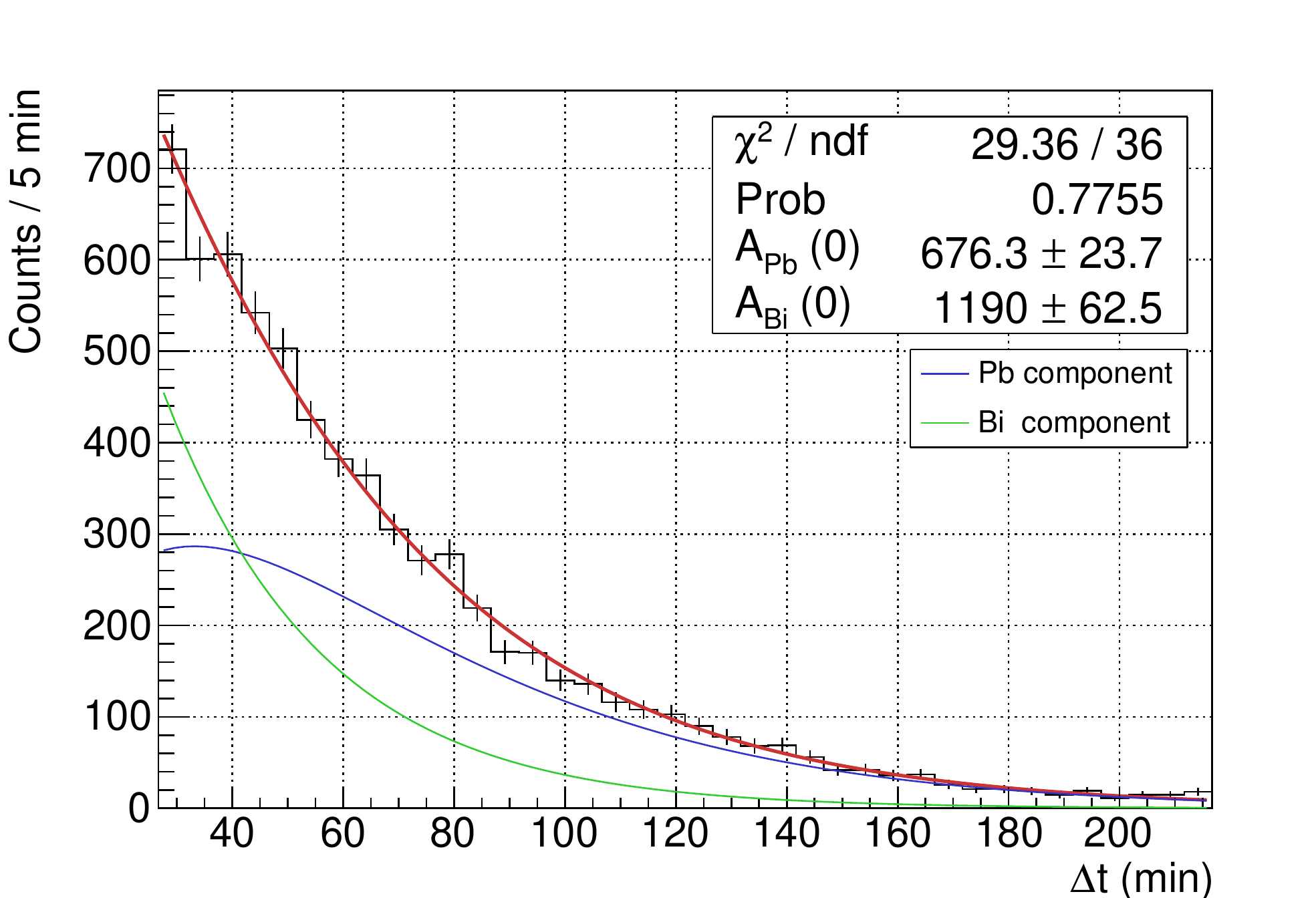}
    \caption{Event rate of \biduq$-$\poduq coincidences versus time. We use the Bateman equation solved for the \pbduq$\rightarrow$\biduq$\rightarrow$\poduq sub-chain to interpolate the experimental data. The two summed terms in the Bateman equation solution, corresponding to the decays of \pbduq and \biduq nuclei initially present in the sample, are represented with blue and green functions, respectively.}
    \label{fig:exp_U238}
\end{figure}

We analyze the data to search for the \biduq$-$\poduq delayed coincidences. Our detector amplifier, being characterized by a shaping time of the order of 1~$\mu$s, allows to separately record the energies of \biduq and \poduq events. 
However, since the \poduq half-life is shorter than the time resolution of our acquisition system (limited by the MCA timer frequency), the same timestamp is assigned to both events. 
In Fig.~\ref{fig:Spectrum_U238}, we show the measured spectrum and we highlight the \biduq$-$\poduq correlated events satisfying the following requirements:
\begin{itemize}
    \item the \biduq events must have an energy below the Q-value of the $\beta$ decay (3.3~MeV);
    \item the $^{214}$Po events must be recorded at the 7.687~MeV $\alpha$ peak (in a range extending from 7.42 to 7.72~MeV) within 1~$\mu$s from the \biduq event.
\end{itemize}
In order to verify the goodness of our selection, we analyze the \biduq$-$\poduq event rate ($R_{Bi-Po}$) as a function of time. For this purpose, we consider only the data collected after the \poduo has almost completely decayed, discarding the initial part of the measurement corresponding to 5 half-lives of \poduo ($\tau_{1/2}=3.1$~min). 
In this way, by solving the Bateman equations for the \pbduq$\rightarrow$\biduq$\rightarrow$\poduq decay sub-chain, we can
fit the following function to the selected \biduq$-$\poduq data:

\begin{equation}
    R_{Bi-Po}(t)= \frac{A_{Pb}(0) \lambda_{Bi}}{\lambda_{Bi}-\lambda_{Pb}} \left(e^{-\lambda_{Pb}t} - e^{-\lambda_{Bi}t} \right) + A_{Bi}(0) e^{-\lambda_{Bi}t}
\label{eq:bateman}
\end{equation}
where $A_{Pb}(0)$ and $A_{Bi}(0)$ are, except for an efficiency term, the activities of \pbduq and \biduq at the beginning of the considered data taking, whereas $\lambda_{Pb}$ and $\lambda_{Bi}$ are their decay constants. 
In order not to have too many free parameters in the fit, we fix the decay constants to the values reported in literature~\cite{WU2009681} and we obtain the result shown in Fig.~\ref{fig:exp_U238}, where we highlight the contributions from the two summed terms in Eq.~\ref{eq:bateman} due to \pbduq and \biduq decays, respectively. The chi-square test demonstrates that the experimental data are well interpolated, thus proving that the event rate as a function of time of \biduq$-$\poduq selected data is consistent with the expectations.

To complete the analysis of this experimental measurement we simulate it with the MC model of the set-up, including the detector and the copper plate.
We model the radioactive source, produced by implanting \rnddd daughters, as a surface contamination with an exponential density profile $e^{-x/ \lambda}$, where $\lambda$ is the average implantation depth.
We tune the $\lambda$ parameter to match the line shape of the \poduq $\alpha$ peak, obtaining a value of $\lambda = 52 \pm 1$~nm.
Then, we search for \biduq$-$\poduq coincidences in the MC data, applying the same procedure used for the experimental data.
By analyzing the $\beta$-spectrum shape and integral of the selected \biduq events, we find (by scanning different thicknesses) that the active layer to be used in MC simulations to match the experimental data is 400$\pm$50~$\mu$m.
In Fig.~\ref{fig:Spectrum_U238} we show the MC spectrum obtained by simulating \biduq$-$\poduq decays in comparison with the experimental data.
The MC spectrum is normalized to the counts recorded at the \poduq $\alpha$ peak and, as expected, does not fully reproduce the experimental data at low energies because of the contributions from other $\beta$ decays and from the environmental background.
By looking at the selected spectra of delayed coincidences, we observe a good agreement (within 10\%) between the experimental data and the MC reconstruction both at the \poduq $\alpha$ peak and in the energy range of \biduq $\beta$ emission. 
This is a remarkable achievement because it highlights our capability to select the \biduq events in the low energy spectrum that includes contributions from other background sources.

\subsection{\textbf{Analysis of \raddt$-$\rndun$-$\poduc delayed coincidence (\udtc chain)}}

\begin{figure}[t]
    \centering
    \includegraphics[width=0.49\textwidth]{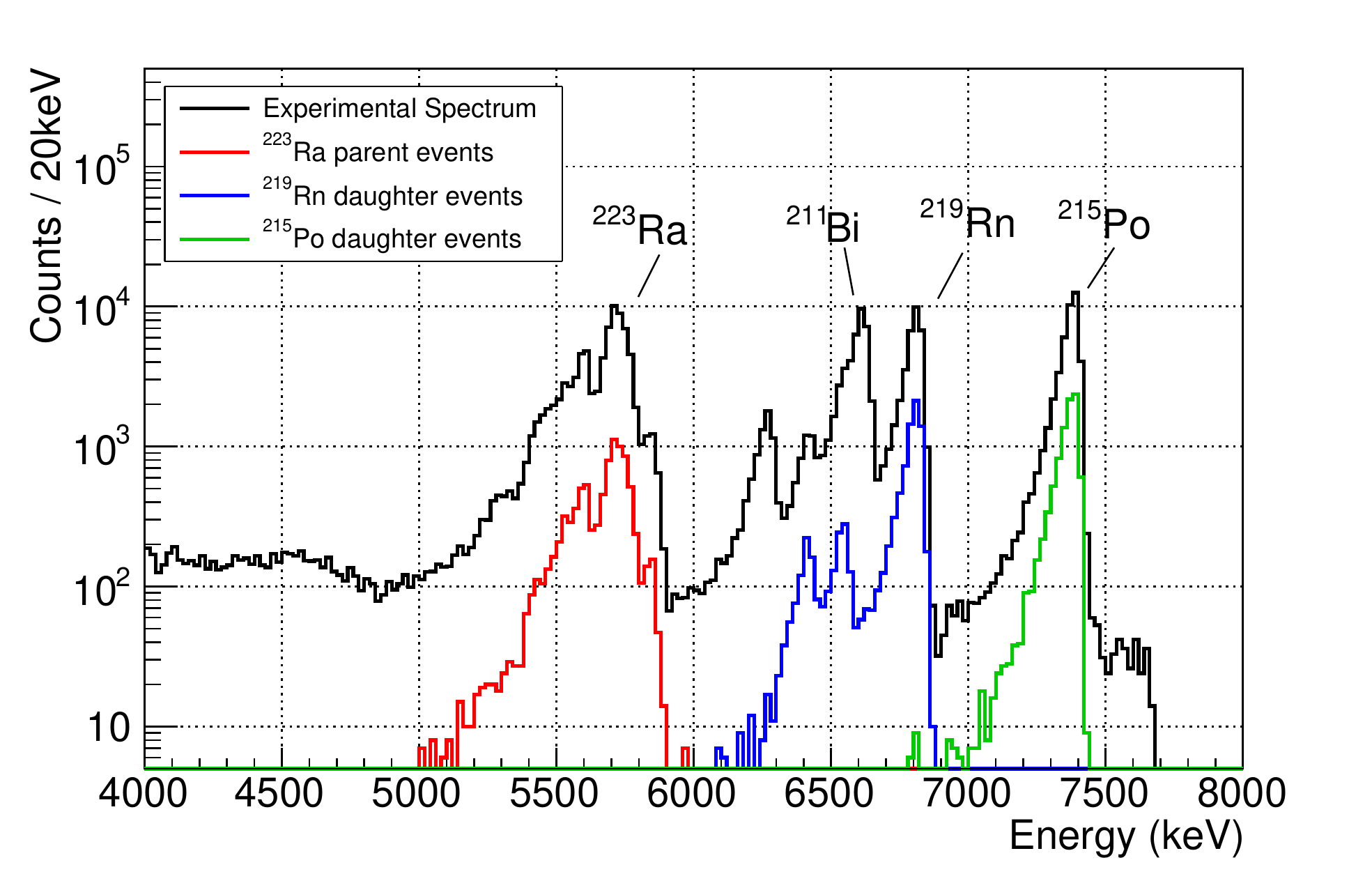}
    \caption{Experimental spectrum (black) and selection of \raddt parent events (red) followed by \rndun and \poduc daughter events (blue and green, respectively).}
    \label{fig:Spectrum_U235}
\end{figure}

\begin{figure}[b]
    \centering
    \includegraphics[width=0.49\textwidth]{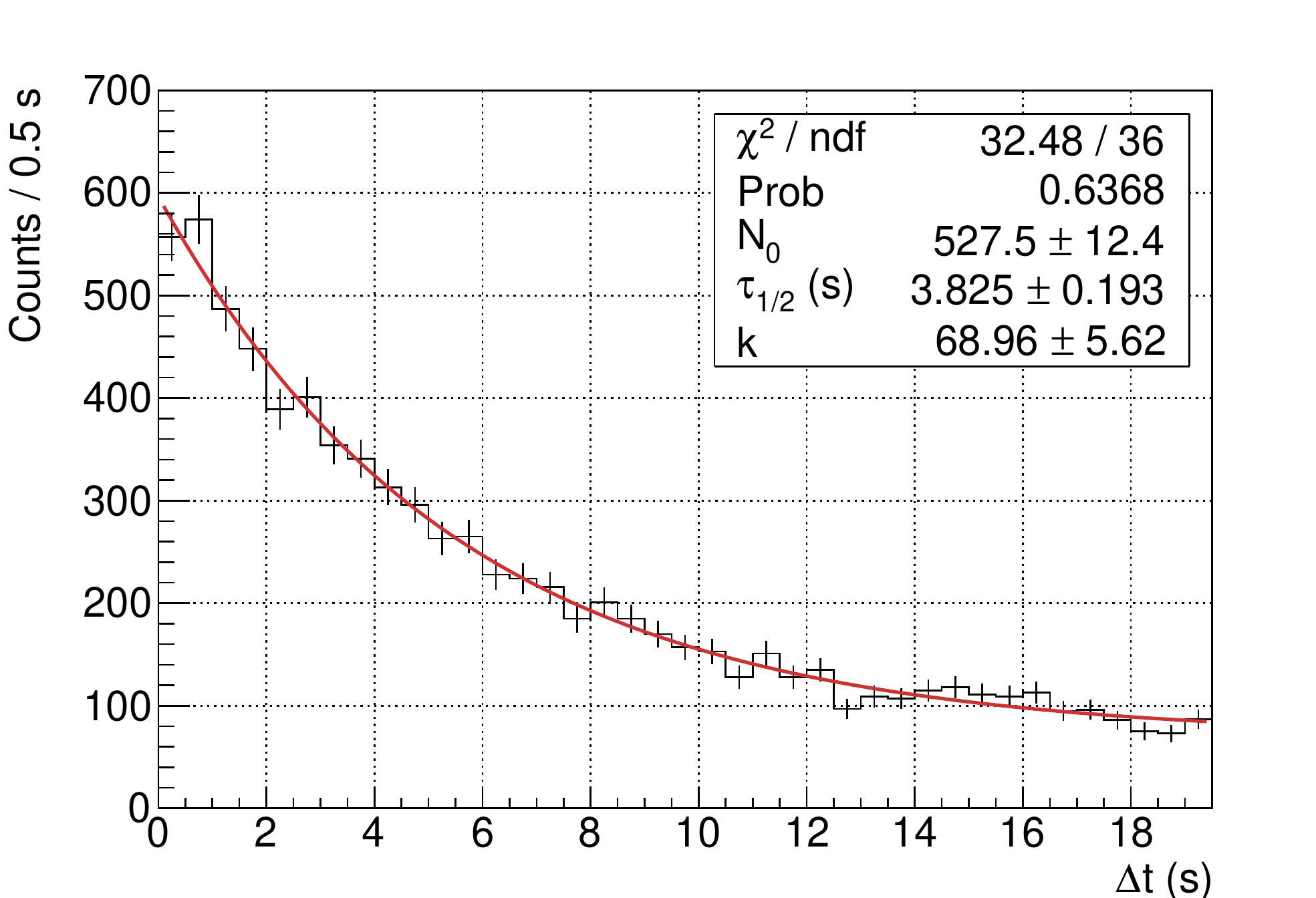}
    \caption{Distribution of the time intervals between \raddt--\rndun decays, interpolated with an exponential function plus a flat background to account for random delayed coincidences.}
    \label{fig:exp_U235}
\end{figure}

The second test is performed with a source of \raddt, belonging to the \udtc decay chain. The radioactive sample consists of a $1.5 \times 1.5 \times 0.05$~cm$^3$ copper plate that was exposed to a $^{227}$Ac source, thus implanting \raddt and its daughters.
We measured the copper plate for 53.2 days using the same experimental set-up previously described.
The measured spectrum, shown in Fig.~\ref{fig:Spectrum_U235}, exhibits prominent peaks at the energies of the $\alpha$-particles emitted by the source.
From the analysis of $\alpha$-peak intensities, we observe a secular equilibrium break at \rndun decay, whose average activity is nearly half that of \raddt.
Nevertheless, we can analyze the collected data to search for subsequent decays of \raddt$-$\rndun$-$\poduc, characterized by 11.4~d, 3.96~s, and 1.78~ms half-lives, respectively.
We use the following criteria to select sequences of three consecutive $\alpha$-decays:
\begin{itemize}
    \item the \raddt parent events must fall in the [5$-$6]~MeV energy range, which includes the main $\alpha$ lines of this decay;
    \item the \rndun daughter events in the [6$-$7]~MeV range must be recorded within a time window $\Delta t_w = 19.8$~s (corresponding to $5 \, \tau_{1/2}$ of \rndun) and must be followed by a \poduc event detected within a very short time (17.8~ms).
\end{itemize}
The coincident events are highlighted in Fig.~\ref{fig:Spectrum_U235}, where it is possible to appreciate that the spectrum of daughter events is dominated by the peaks at the energies of the $\alpha$-particles emitted by \rndun and \poduc.
Since in this case the half-life of the daughter nucleus \rndun is much longer than the time resolution of the system, we can study the distribution of time intervals ($\Delta t$) between couples of \raddt$-$\rndun events.
In Fig.~\ref{fig:exp_U235}, we fit to the experimental $\Delta t$ data a function of the form:

\begin{equation}
     N(\Delta t) = N_0 \; 2^{- \Delta t / \tau_{1/2}} + k
\label{eq:deltaTfit}
\end{equation}
where $\tau_{1/2}$ is a free parameter which is expected to be compatible with the \rndun half-life, $N_0$ is the amplitude of the exponential term, and $k$ is a flat component to account for random delayed coincidences in the approximation that their rate ($r_R$, which is $\mathcal{O}(10^{-3}$)~Hz at most) satisfies the condition $r_R\,\Delta t_w \ll 1$.
The fit results are compatible with the expectations, confirming the effectiveness of the selection technique.
Particularly, the value determined for the \rndun half-life ($\tau_{1/2} = 3.83 \pm 0.19$~s) is in perfect agreement with that reported in literature~\cite{MAPLES1977223}. 

The measurement of delayed coincidences allows us to evaluate the detection efficiency of $\alpha$-particles directly from the experimental data.
We can reasonably assume that the detection efficiency of $\alpha$-particles ($\varepsilon_{\alpha}$) is nearly constant in the energy range considered for this analysis. This is also confirmed by our MC simulations.
Since \rndun and \poduc have the same activity, the following equations hold:

\begin{equation}
\begin{cases}
    N_{Po} = N_{dec} \, \varepsilon_{\alpha}\\
    N_C = N_{dec} \, \varepsilon_{\alpha}^2 \, f
\end{cases}
\label{Eq:system}
\end{equation}

In this system $N_{Po}$ is the total number of events recorded at the \poduc $\alpha$-peak, $N_{dec}$ are the \rndun (or equivalently \poduc) decays occurred during the measurement, $N_C$ are the \rndun$-$\poduc delayed coincidences, and $f$ is the fraction of decays expected to occur in the time window set to search for \poduc daughters. 
By solving the system in Eq.~\ref{Eq:system}, we get that:

\begin{equation}
\varepsilon_{\alpha} = \frac{N_C}{N_{Po}\,f}
\label{Eq:effAlfa}
\end{equation}

The search for \rndun$-$\poduc delayed coincidences, i.e. events in the [6$-$7]~MeV range followed by a \poduc $\alpha$ within 17.8~ms, gives us $N_C=14268 \pm 98$ (Binomial uncertainty). Thanks to the strong constraint provided by the relatively short half-life of \poduc, the expected number of random delayed coincidences is negligible.
To determine $N_{Po}$, we integrate the counts in the [6.9$-$7.5]~MeV range (black spectrum in Fig.~\ref{fig:exp_U235}), obtaining $N_{Po}=44293$. Since \poduc is the only $\alpha$-emitter in this energy range, the background from other possible $\alpha$ sources is negligible.
Finally, since we open a time window of 17.8~ms, corresponding to 10 half-lives of \poduc, we can set $f=1$. 
The $\alpha$-detection efficiency determined through Eq.~\ref{Eq:effAlfa} is $\varepsilon_{\alpha}=(32.2 \pm 0.2)\%$. 
The efficiency evaluated from the MC simulations analyzing the \rndun and \poduc peaks results is $\varepsilon_{\alpha}^{MC} = (30.3 \pm 0.3)\%$.
The comparison between the experimental and the MC reconstruction of the $\alpha$-detection efficiency gives us a difference which cannot be traced back to a statistical fluctuation, thus pointing out a systematic uncertainty of the order of 5\%.

\subsection{\textbf{Analysis of \rnddz$-$\podus delayed coincidence (\th chain)}}

\begin{figure}[b]
    \centering
    \includegraphics[width=0.49\textwidth]{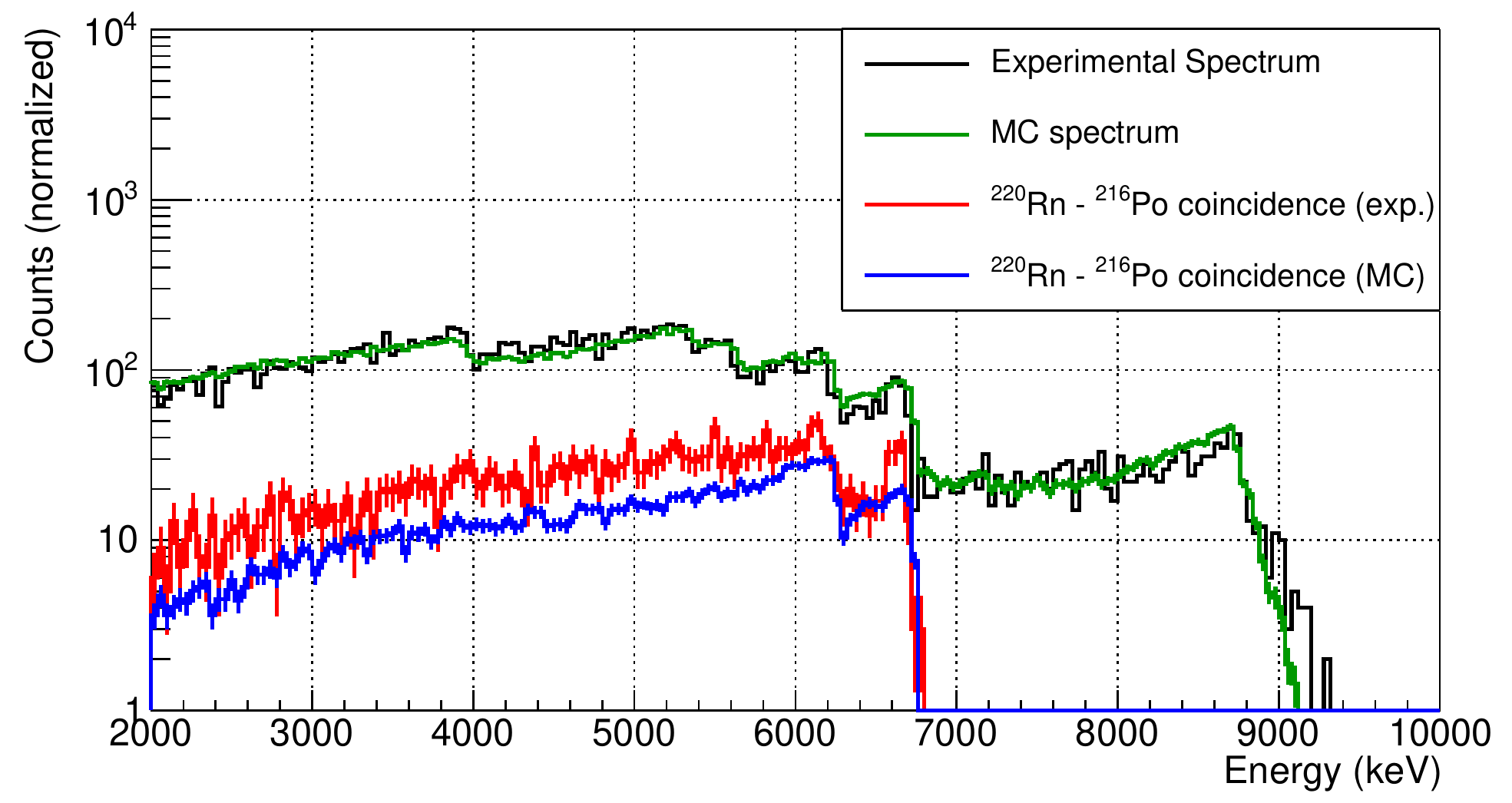}
    \caption{Measured spectrum (black) and selection of \rnddz--\podus delayed coincident events (red). We show the MC simulated spectrum of the full \th chain (green) normalized to the same integral of the measured spectrum in the $E>2$~MeV range, and the corresponding selection of \rnddz--\podus coincidences (blue). The difference between the selected coincidences in the experimental data with respect to the MC simulations is compatible with the number of random coincidences estimated from the plateau in the fit of Fig.~\ref{fig:exp_Th232}.}
    \label{fig:Spectrum_Th232}
\end{figure}

\begin{figure}[t]
    \centering
    \includegraphics[width=0.49\textwidth]{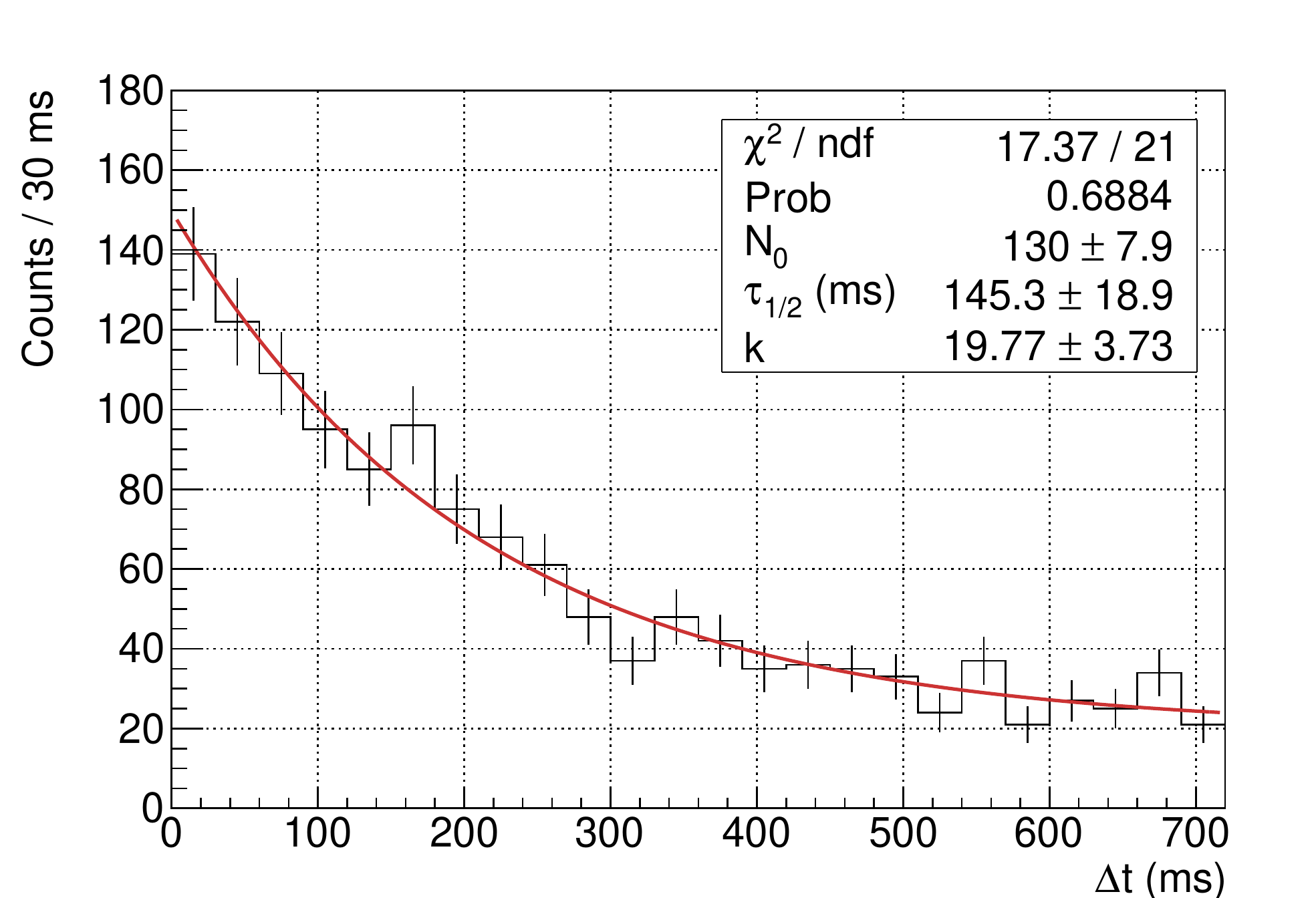}
    \caption{Distribution of the time intervals between \rnddz$-$\podus decays, interpolated with an exponential function plus a flat background to account for random delayed coincidences.}
    \label{fig:exp_Th232}
\end{figure}

To perform the third test, we produced a source by pouring and then evaporating a diluted \th liquid standard on a $5\times5\times0.1$~cm$^{3}$ copper support. 
Assuming secular equilibrium, we measured the sample activity to be $32 \pm 6$~mBq through a $\gamma$-spectroscopy measurement with a low background High Purity Germanium detector.
Then, we measured the \th source with the $\alpha$-spectroscopy apparatus for 73 hours.
Differently from the previous measurements, the measured spectrum (Fig.~\ref{fig:Spectrum_Th232}) does not show any $\alpha$ peak.
We trace this feature back to the formation of an oxidized layer due to the reaction between the nitric acid in the \th solution and the copper. 
This layer of copper oxide was clearly visible on the sample surface and, in light of the experimental measurement, it is thick enough to degrade in energy the emitted $\alpha$-particles.

In this measurement, we search for \rnddz$-$\podus delayed coincidences using the same experimental setup. The challenge we have to face in this case is that there are no clear energy signatures to distinguish the different $\alpha$s emitted by the \th source. Nevertheless, we can rely on the time information, which becomes fundamental.

We set a very wide energy window, from 2~MeV to the maximum energies of \rnddz and \podus $\alpha$s (6.3~MeV and 6.8~MeV, respectively), to select parent-daughter coincidences within a $\Delta t_w = 725$~ms, corresponding to 5 half-lives of \podus ($\tau_{1/2}=145$~ms). 
In Fig.~\ref{fig:Spectrum_Th232}, we show the spectrum of parent-daughter selected coincidences.

To validate our selection, we fit the distribution of the time intervals between \rnddz$-$\podus events with the function of Eq.~\ref{eq:deltaTfit}  (Fig.\ref{fig:exp_Th232}), obtaining a decay constant perfectly compatible with the \podus half-life~\cite{WU20071057,Beretta:2021}. 
Also in this case the rate of random coincidences ($r_R \sim \mathcal{O}(10^{-3}$)~Hz) is small enough to satisfy the condition $r_R\,\Delta t_w \ll 1$, which justifies our choice to fit the background with a flat function.

Next, we exploit MC simulations to describe the source-detector system and extract information about the detection efficiency. 
In order to simulate energy-degraded $\alpha$s, we use a depth parameter of the order of the range of $\alpha$-particles in copper (which we assume to be similar to the one in copper oxide). 
Particularly, we choose $\lambda=5~\mu$m, which gives us a good reconstruction of the spectral shape in the whole range of energy-degraded $\alpha$s (Fig.~\ref{fig:Spectrum_Th232}, green spectrum).
By selecting the delayed coincidences in the MC simulation, we get that in this measurement (with a point-like source faced to the detector) the probability to observe a \rnddz$-$\podus delayed coincidence involving two events with $E>2$~MeV is $(7.3\pm0.4)$\%, including a 5\% systematic uncertainty.
After subtracting the random component evaluated through the fit in Fig.~\ref{fig:exp_Th232}, the experimental number of delayed coincidence is $897\pm97$.
Thus, from the analysis of delayed coincidences, we get that the sample activity is $47\pm5$~mBq, which is statistically compatible at 2$\sigma$ with the one determined through the aforementioned $\gamma$-spectroscopy measurement ($32\pm6$~mBq).

This result proves that even in the absence of clear signatures in the energy spectrum, we are able to identify a relatively deep surface contamination and evaluate its activity thank to the analysis of time correlations.
As shown in the next section, this approach is particularly useful in the analysis of samples with a low contamination level.

\section{Analysis of a low activity sample} \label{Sec:low}
\begin{figure}[b]
    \centering
    \includegraphics[width=0.49\textwidth]{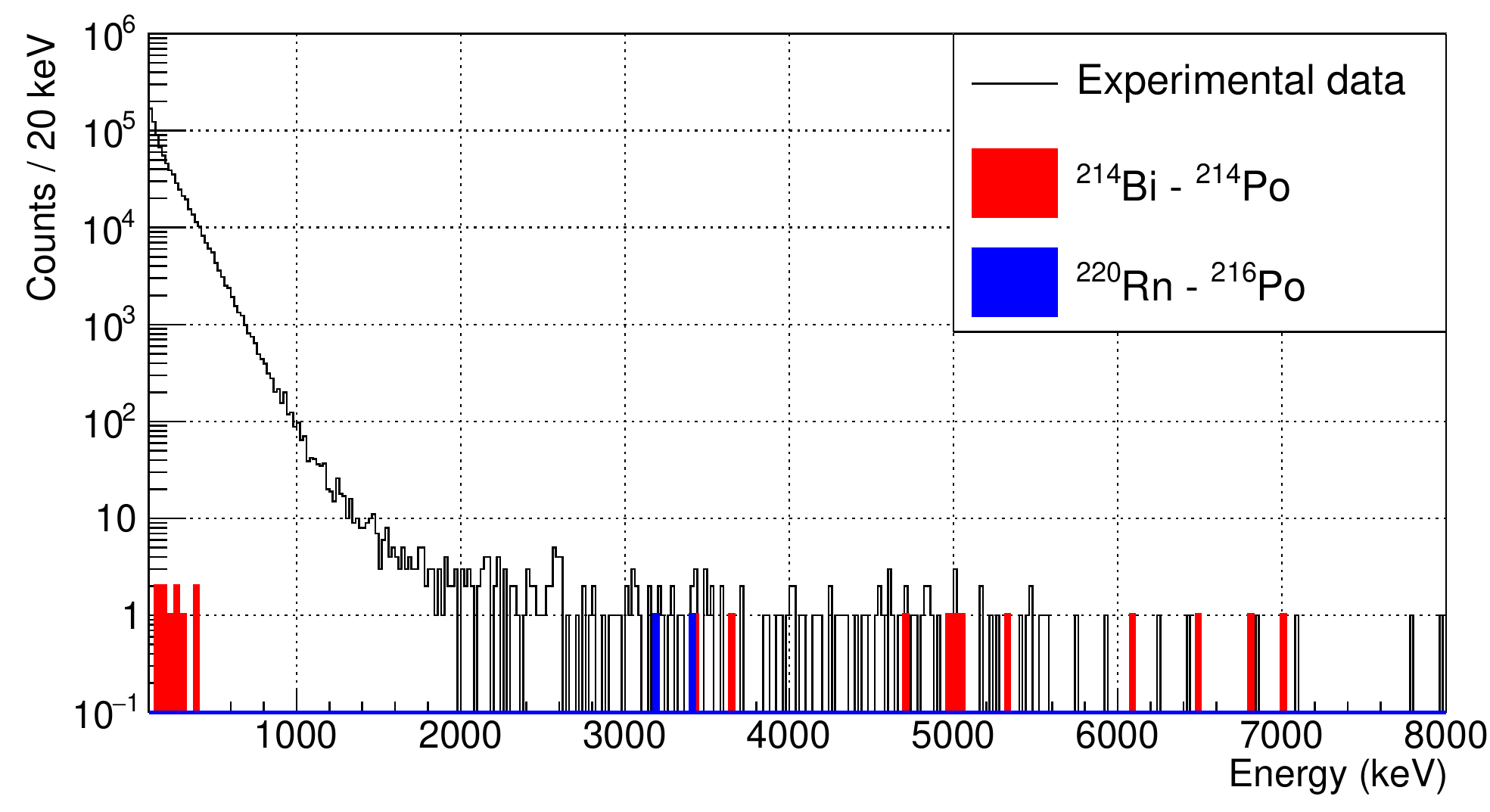}
    \caption{Measured spectrum (black) and selection of \biduq$-$\poduq (red) and \rnddz$-$\podus (blue), belonging to the \u and \th decay chains, respectively. The analysis of time correlations allows to disentangle the \u and \th contamination in the continuum produced by energy-degraded $\alpha$s.}
    \label{fig:sensitivity}
\end{figure}

After testing and validating the selection of delayed coincidences with different sources, we apply this technique to analyze a weakly contaminated sample. This allows to probe the sensitivity of the measurement apparatus and to highlight the potential of this method in the framework of low background measurements.
The weakly radioactive sample consists of a $5\times5\times0.1$~cm$^{3}$ high-purity copper plate (OFHC, Oxygen-Free High Conductivity) treated with surface plasma ablation, a typical cleaning technique for materials used in rare event searches~\cite{Azzolini:2018tum}.
In Fig.~\ref{fig:sensitivity}, we show the spectrum collected with a 31.9 days measurement. In the $\alpha$ region above the 2.6 MeV $^{208}$Tl $\gamma$-line, we do not observe any peak, thus the analysis of the energy spectrum does not allow to identify any particular $\alpha$ emitter.
Nevertheless, as shown in the previous section, the analysis of time correlations is a powerful tool to increase our knowledge about contaminations even in the absence of $\alpha$ peaks.
Therefore, we analyze the data to search for delayed coincidences involving the energy-degraded $\alpha$ events that populate the spectrum above the 2615~keV $^{208}$Tl line (representing the endpoint of the main environmental $\beta/\gamma$ radiation). 
Taking into account the energy resolution of the detector, we set at $E=2.65$~MeV the lower bound for the energy region of degraded $\alpha$s. 
The total number of events detected in this energy region is 122, thus the rate of degraded $\alpha$s is $r_{\alpha}=(4.4 \pm 0.4)\times10^{-5}$~Hz.

The data analysis strategy is set up in the following steps with the goal of identifying possible contaminants belonging to the \th and \u decay chains (\udtc contaminations can be neglected due to its low isotopic abundance in nature).

\begin{enumerate}
    \item In order to investigate \u chain contaminations, we search for \biduq$-$\poduq coincident events between a $\beta$ with $0.1<E<2.65$~MeV and an energy-degraded $\alpha$.
    In this search, there is no ambiguity with the \bidud$-$\podud sequence of the \th chain, because the \podud half-life ($\tau_{1/2}=0.3$~$\mu$s) is shorter than the shaping time of the amplifier (1~$\mu$s) and, thus, it produces piled-up events.
    We find 11 coincidences in the collected data. 
    Taking into account that we opened $8.73\times10^5$ time-windows with $\Delta t_w = 1$~ms, the expected number of random coincidences is $\sim 0.04$. 
    Therefore, we can reasonably state that all the 11 observed coincidences are produced by a \biduq$-$\poduq decay sequence belonging to the \u chain.
    \item To probe \th chain contaminations, we search for \rnddz$-$\podus delayed coincidences between degraded $\alpha$s occurring within a $\Delta t_w = 0.725$~s. We find only 1 pair of events satisfying this requirement, in front of an expected number of random coincidences estimated to be $\sim 0.004$. Since there are no other decay sequences occurring on this time scale in both \th and \u chains, this event can be attributed to a \rnddz$-$\podus cascade.
    \item As a final cross-check, we search for delayed coincidences between degraded $\alpha$s with $0.725$~s $< \Delta t$ $< 9.3$~min, (upper bound corresponding to $3\times\tau_{1/2}$ of \poduo). Indeed, since we observed 11 \biduq$-$\poduq events, we expect to detect also \rnddd$-$\poduo sequences belonging to the previous part of the \u chain. 
    The selection criteria adopted in this case include also the \raddq$-$\rnddz ($\tau_{1/2}=55.6$~s) sequence belonging to the \th chain. However, given the results obtained from the previous selections, the dominant contribution is expected to come from the \u chain.
    In this case we find 7 delayed coincidences with an expected number of random $\sim 2$.
    This result is compatible with the previous ones, because, according to our MC simulations, the efficiency to detect a $\beta - \alpha$ \biduq$-$\poduq coincidence is a factor $\sim 2.1$ higher with respect to the $\alpha - \alpha$ \rnddd$-$\poduo one.
\end{enumerate}

In Table~\ref{tab:results}, we summarize the results obtained from this analysis. 
Since we found delayed coincidences involving energy-degraded $\alpha$s, the contamination depth must be on the same scale of the $\alpha$-range in copper (i.e few tens of $\mu$m). 
Thus, in order to evaluate the efficiencies to detect a delayed coincidence ($\varepsilon_C$), we simulate the surface contamination of the sample with a depth parameter $\lambda = 10~\mu$m. 
The efficiencies reported in Table~\ref{tab:results} are obtained by generating the decays on the whole sample surface (52~cm$^2$), in front of a detector active area of 9~cm$^2$.
We obtain contaminant activities per unit of surface area ($A_s$) on the order of $\mu$Bq/cm$^2$ for both \u and \th chains. 
Given the relatively small number of detected coincidences, we apply the Feldman-Cousins approach~\cite{PhysRevD.57.3873} to calculate the 90\% C.L. interval for known mean background equal to zero and to assess the uncertainty range on $A_s$ results. 

\begin{table}[t]
    \centering
    \caption{Summary of the analysis of delayed coincidences in the measurement of a low activity copper sample. We list the selection criteria and the results in the search for \biduq$-$\poduq and \rnddz$-$\podus delayed coincidences. The number of detected coincidences ($N_C$) are finally combined with absolute efficiencies ($\varepsilon_C$) to extract the contaminant activities per surface unit ($A_s$). The uncertainties on $A_s$ correspond to the 90\% C.L. interval evaluated through the Feldman-Cousin statistical approach.}
    \begin{tabular}{l|c|c}
    Decay chain & \u & \th\\
    \hline
    Decay sequence & \biduq$-$\poduq & \rnddz$-$\podus \\
    Decay type & $\beta - \alpha$ & $\alpha - \alpha$ \\
    Range parent (MeV) & [0.1 -- 2.65] & [2.65 -- 6.32] \\
    Range daughter (MeV) & [2.65 -- 7.72] & [2.65 -- 6.81] \\
    $\Delta t_w$ (ms) & 1 & 725 \\
    $N_C$ (Delayed coincidences) & 11 & 1 \\
    $\varepsilon_C$ (\%) & 1.10 $\pm$ 0.01 &  0.74 $\pm$ 0.01\\
    $A_s$ ($\mu$Bq/cm$^2$) & 7.6$_{-3.2}^{+4.3}$ & 0.9$_{-0.8}^{+3.2}$ \\
    \end{tabular}
    \label{tab:results}
\end{table}

The sensitivity we obtain through the analysis of delayed coincidences is on the same scale as that which would be obtained assuming that the events observed in the higher energy region of the spectrum ($E>6$~MeV) are produced by $\alpha$ decays belonging to the \th chain (\podud, $E_{\alpha}=8.785$~MeV) or to the \u one (\poduq, $E_{\alpha}=7.687$~MeV).
Particularly, there are 3 events with $E>7.72$~MeV (that can not originate from the \u chain) and 8 events in the range from 6 to 7.72~MeV, which can be produced by both decay chains. If we conservatively attribute the counts in these ranges to \th and \u chains, respectively, we would obtain that the activities are at maximum $2.5_{-1.6}^{+3.6}$~$\mu$Bq/cm$^2$ for \th and $3.4_{-1.7}^{+2.6}$~$\mu$Bq/cm$^2$ for \u (which are compatible with those reported in Tab.~\ref{tab:results} taking into account their respective 90\% confidence intervals).
In both cases, we obtain results on the $\mu$Bq/cm$^2$ sensitivity scale, because the efficiency loss due to the search of a coincident event is practically fully compensated by the fact that we can extend the energy range down to 2.65~MeV.
However, thanks to the analysis of delayed coincidences, we get clear signatures to disentangle the \u chain contaminations from the \th ones, allowing to determine their activity without ambiguity in the identification of the radioactive contaminant (\th or \u chain) that produces the energy-degraded $\alpha$s. 
Indeed, the discrimination power of this technique to identify \u versus \th contaminations, evaluated as the probability that a \rnddz$-$\podus coincidence happens with a $\Delta t>1$~ms and, thus, is not tagged as a \biduq$-$\poduq one is $>99.5\%$.
For example, in this test case, we have been able to assess that the \u contamination activity is higher than the \th one, and such information can be extremely useful in the framework of radiopurity assay measurements. 
Moreover, the observation of delayed coincidences involving energy-degraded $\alpha$s, allows us also to extract precious information about the features of surface contaminations in terms of average contamination depth.

Finally, as can be seen in Fig.~\ref{fig:sensitivity}, there are no delayed coincidences detected in the ranges where \poduq, \rnddz, and \podus $\alpha$-peaks would be expected to show up in the case of very shallow surface contaminations. 
In order to provide a limit for such kind of contaminations, we run the MC simulations with a depth parameter $\lambda = 10$~nm small enough to make the $\alpha$-energy degradation effect negligible.
By evaluating the efficiencies to detect delayed coincidences at the $\alpha$-peaks (in a $\pm$~FWHM range), and calculating 90\% C.L. upper limits, we obtain the following results:

\begin{equation*}
    \begin{split}
        A_s (^{214}\text{Bi} - ^{214}\text{Po}) < 0.79~ \mu\text{Bq/cm}^2~~(90\%~\text{C.L.})\\
        A_s (^{220}\text{Rn} - ^{216}\text{Po}) < 0.76~ \mu\text{Bq/cm}^2~~(90\%~\text{C.L.})   
    \end{split}
\end{equation*}
for the \u and \th chain, respectively.

\section{Conclusions}
In this article, we described the delayed coincidence analysis technique and its application to $\alpha$-spectroscopy measurements with silicon detectors.
We validated the data acquisition software and the analysis procedure through dedicated measurements with different contaminated samples, demonstrating the reliability and the flexibility of this technique. Moreover, we highlighted the potential of this analysis applying it to the radiopurity assessment of an ultra-pure sample.
Although the achieved sensitivity is comparable to the one calculated with the standard approach ($\sim\mu$Bq/cm$^{2}$), the time correlation allowed us to disentangle contaminations of \u chain from those of \th one with a discrimination power $>99.5\%$, determining their activity and average implantation depth.
This information is essential for several reasons, such as the investigation of contamination mechanisms (e.g. deposition, implantation, diffusion) and the development of new purification strategies for further improvement of material radiopurity.
These aspects will play a crucial role in the selection of materials for the next generation experiments~\cite{CUPIDInterestGroup:2019inu}. Finally, the analysis of delayed coincidences can also be applied to identify and reject background events from surface contaminations of detector materials.
\begin{acknowledgements}
This work makes use of the Arby software for Geant4 based Monte Carlo simulations, that has been developed in the framework of the Milano -- Bicocca R\&D activities and that is maintained by O. Cremonesi and S. Pozzi.
\end{acknowledgements}

\bibliography{bibliography}  
\bibliographystyle{spphys} 
\end{document}